\documentclass[iop]{emulateapj}

\usepackage{bm}

\slugcomment{Draft for ApJ}

\shorttitle{Photodesorption and Surface Snow Line in Protoplanetary Disks}
\shortauthors{Oka et al.}

\begin{document}


\title{Effect of Photodesorption on Snow Line at the Surface of 
Optically Thick Circumstellar Disks around Herbig Ae/Be Stars}


\author{Akinori Oka}
\affil{Department of Earth and Planetary Sciences, Tokyo Institute of
Technology, Ookayama, Meguro, Tokyo 152-8551, Japan}
%\email{akinori@geo.titech.ac.jp}

\author{Akio K. Inoue}
\affil{College of General Education, Osaka Sangyo University, 3-1-1,
Nakagaito, Daito, Osaka 574-8530, Japan}
\email{akinoue@las.osaka-sandai.ac.jp}

\author{Taishi Nakamoto}
\affil{Department of Earth and Planetary Sciences, Tokyo Institute of
Technology, Ookayama, Meguro, Tokyo 152-8551, Japan}

\and

\author{Mitsuhiko Honda}
\affil{Department of Information Science, Kanagawa University, 2946, 
Tsuchiya, Hiratsuka, Kanagawa 259-1293, Japan}


%\altaffiltext{1}{Department of Earth and Planetary Sciences, Tokyo Institute of Technology}

%\altaffiltext{2}{College of General Education, Osaka Sangyo University}

%\altaffiltext{3}{Department of Information Science, Kanagawa University}


\begin{abstract}
We investigate the effect of photodesorption on the snow line position
 at the surface of a protoplanetary disk around a Herbig Ae/Be star,
 motivated by the detection of water ice particles at the surface of the
 disk around HD142527 by Honda et al. For this aim, we obtain the
 density and temperature structure in the disk with a 1+1D radiative
 transfer and determine the distribution of water ice particles in the
 disk by the balance between condensation, sublimation, and
 photodesorption. We find that photodesorption induced by the
 far-ultraviolet radiation from the central star depresses the
 ice-condensation front toward the mid-plane and pushes the surface snow
 line outward significantly when the stellar effective temperature
 exceeds a certain critical value. This critical effective
 temperature depends on the stellar luminosity and mass, the water
 abundance in the disk, and the yield of photodesorption. We present an
 approximate analytic formula for the critical temperature. We separate
 Herbig Ae/Be stars into two groups on the HR diagram according to the
 critical temperature; one is the disks where photodesorption is
 effective and from which we may not find ice particles at the surface,
 and the other is the disks where photodesorption is not effective. We
 estimate the snow line position at the surface of the disk around
 HD142527 to be 100--300 AU which is consistent with the water ice
 detection at $>140$ AU in the disk. All results depend on the dust
 grain size by a complex way and this point requires more work in
 future.
\end{abstract}





\section{Introduction}

In a protoplanetary disk, there is the so-called snow line which is
defined as the inner boundary of the ice-condensing region
\citep[e.g.,][]{hayashi81}. Beyond the snow line, the surface mass
density of solid materials is enhanced due to condensation of water ice,
and hence, the formation efficiency of large gaseous and icy planets is
expected to be enhanced. Thus, the snow line plays a very important role
in the planet formation.

In spite of its importance, no one has detected the snow line position
in a circumstellar disk so far. This is because the snow line is usually
located close to the central star ($R\lesssim 10~\mathrm{AU}$), where the
spacial resolving power of current facilities is limited, and the
disk is optically too thick to see through. On the other hand, the
presence of water ice in circumstellar disks have already been detected
\citep[e.g.,][]{malfait99,terada07,honda09}. \cite{terada07} observed
scattered radiation from an {\it edge-on} disk and detected an
absorption feature of water ice at 3 \micron\ in it. However, their
observations give no information about the position of the snow line
along the line of sight. \cite{honda09} observed scattered light from
the surface of an {\it face-on} disk and detected the $3~\mathrm{\mu m}$
absorption feature by icy dust particles at $R\gtrsim 140~\mathrm{AU}$. 
They could estimate the dust size as $\sim1$ \micron\
by using a color diagram method proposed by \cite{inoue08}. While
\cite{honda09} could not detect the snow line position, their method
holds a potential to achieve it due to a high spatial resolution of
their detection technique.

Although an important factor for the planet formation is the snow
line position at the mid-plane, detecting the snow line at the disk
surface will be a first step because the disk surface is easily
observed. The surface snow line is connected to the mid-plane snow line
through a 2D structure of the ice-condensing region
\cite[e.g.,][]{davis05,oka11}. If one knows the mechanism which
determines the 2D structure of the snow line, one may convert the
surface snow line into the mid-plane snow line.

The main difference between the interior and the surface of
protoplanetary disks is that the latter is exposed to ultra-violet (UV)
radiation from the central star and the interstellar radiation
field. This UV radiation may affect the distribution of water ice at the
disk surface. Indeed, it has been proposed that a non-thermal desorption
process is necessary for explaining the observed high abundance of
gas-phase water molecules in cold outer regions of circumstellar disks
and molecular clouds \citep[e.g.,][]{dominik05,hollenbach09}. A probable
mechanism responsible for such a non-thermal desorption of water
molecules is the photodesorption (or photosputtering) process caused by
far-UV (FUV) radiation. The photodesorption process has been recently
studied theoretically and experimentally
\citep[e.g.,][]{westley95a,westley95b,andersson06,andersson08,oberg09,hama10}.

\"{O}berg et al. (2009) experimentally studied the photodesorption
process and derived the photodesorption yield as a function of ice
temperature and thickness of the ice layer. According to their results,
the photodesorption yield with ice temperature $T$ and thickness of the
ice layer larger than 8 mono layers is  
$Y_{\mathrm{pd}}(T,x>8)=10^{-3}(1.3 + 0.032 T)~{\rm UV~photon^{-1}}$,
where $x$ is the ice thickness scaled by a mono layer. They then carried
out a numerical simulation about the distribution of gas phase
$\mathrm{H_{2}O}$ in a protoplanetary disk around a Herbig Ae/Be star, 
taking into account adsorption and thermal desorption processes on
dust grains, and the photodesorption process, and found that 
photodesorption reduced the H$_2$O ice abundance in the disk surface by
orders of magnitudes.

\cite{grigorieva07} also showed theoretically that the presence
of icy dust particles in a debris disk around a Herbig Ae/Be star is
affected by the photodesorption process. The environment of the surface
of protoplanetary disks such as one observed by \cite{honda09} may be
similar to that of debris disks, and hence the position of the surface
snow line would be affected by the photodesorption process.

Recently, observations of water vapor in protoplanetary disks show
a rapid progress with {\it Spitzer Space Telescope} and {\it Herschel
Space Observatory} as well as large ground-based telescopes 
\citep[e.g.,][]{pon10,stu10,vank10,ber10,fed11}. 
There is lack of hot and warm vapor traced by near- and mid-infrared
observations in disks around Herbig Ae/Be stars, whereas such vapor was
often found in disks around T Tauri stars \citep{pon10,fed11}. A
difference of photochemistry is suggested as a cause of this
difference. In Herbig Ae/Be disks, H$_2$O molecules (photo)desorbed from
icy grains may be photodissociated by strong UV radiation. On the other
hand, \cite{stu10} and \cite{vank10} detected cool water vapor in
far-infrared observations in Herbig Ae/Be disks, suggesting incomplete
photodissociation in the outer part of the disks. More interestingly, 
\cite{ber10} reported no detection of cold water vapor in submillimetre
in a T Tauri disk and suggested that photodesorption was less efficient
than theoretically expected. In any case, more works about
photodesorption and photodissociation in protoplanetary disks are
required to depict a full picture of the photochemical evolution in the
disks. In particular, less efficient photodesorption suggested by
\cite{ber10} should be tested also in Herbig Ae/Be stars.

In this study, we numerically investigate the effect of the
photodesorption process induced by the FUV radiation from the central
star and the interstellar radiation field on the surface snow line
position in circumstellar disks. We focus on Herbig Ae/Be disks
where photodesorption is effective as expected, while the parameter
space explored in this paper also covers T Tauri disks. The goal is
to uncover what the determining process of the snow line position at
a disk surface is. We omit photodissociation in this paper, but in
fact, we find that this process is much slower than photodesorption (see
\S4.3.1).

This paper is organized as follows. 
In \S2, we describe our model for obtaining the distribution of icy dust
particles in a Herbig Ae/Be disk with photodesorption. In \S3, we
show effects of photodesorption on the position of the surface snow
line and the dependence of the results on the stellar parameters. In
\S4, we consider an implication of our results for observations, derive
an analytic formula of the critical stellar effective temperature for an
efficient photodesorption, and discuss effects of some processes
omitted in our calculations. Finally, we summarize this paper in \S5.

\section{Model}

\subsection{Disk structure}

We consider a flared but geometrically-thin and optically-thick disk
revolving around a central star in Kepler velocity. We assume that the
disk is axi-symmetric about the rotational axis and plane-symmetric about
the mid-plane. We adopt a coordinate system $(R,Z)$, where $R$ and $Z$
are the radial distance from the central star and the vertical height
from the disk mid-plane, respectively. The central star sits on the
origin of the coordinate.

We assume that the radial distribution of the surface density of the
disk is a power law distribution \citep[e.g.,][]{hayashi81} as 
\begin{equation}
 \Sigma(R) = \Sigma_{1\mathrm{AU}}
  \left( \frac{R}{1\mathrm{AU}} \right)^{-p},
\end{equation}
where $\Sigma_{1\mathrm{AU}}$ is the surface density at $R=1\mathrm{AU}$
and $p$ is the power-law index describing the radial density
distribution. We assume $\Sigma_{1\mathrm{AU}}=1700\ \mathrm{g~cm^{-2}}$
and $p=1.5$ \citep{hayashi81} throughout this paper. The disk is
assumed to be in the hydrostatic equilibrium vertically, and heated by
the irradiation by the central star. The temperature and density
structures of the disk are numerically solved by the 1+1D radiative
transfer method described by \cite{dul02}. We assume that temperatures of
dust and gas are always same. Unlike \cite{dul02}, we consider isotropic
scattering of both stellar and diffuse radiations according to
\cite{inoue09} and \cite{oka11}. The details of the calculation are
described in Appendix A \citep[see also][]{oka11}. Note that we assume
that the grazing angle ($\beta$) of the stellar radiation incident
on the disk surface is uniformly 0.05 throughout the disk for
simplicity. This is fairly artificial but it was very difficult to
determine both the grazing angle and the ice abundance simultaneously in
our 1+1D disk with scattering by an iteration scheme because of very
slow convergence. The mean molecular weight of the disk gas is assumed
to be 2.3.

Although we fix global parameters of the disk such as 
$\Sigma_{1{\rm AU}}$, $p$, and $\beta$, it would not affect the results
presented in this paper very much when discussing the disk surface where
the radiative transfer effect is negligible and the physical condition
is determined almost locally. This enables us to derive an analytic
formula describing the numerical results in \S4.2, while the
absolute scaling may be affected slightly.

\subsection{Dust opacity}

A very important factor for determining the structure and the radiation
field in the disk is the optical property of the disk. This is mainly
determined by the absorption and scattering processes by dust
particles. The gas opacity is negligible for the radiation
considered in this paper, except for Ly$\alpha$ photons.
We omit gas opacity throughout this paper for simplicity, whereas an
importance of Ly$\alpha$ transfer was suggested \citep{fog11,bet11}. It
is still reasonable when discussing the disk surface where the transfer
effect is small.

We consider only pure crystalline silicate and pure crystalline water
ice particles floating in the disk gas as dust. These particles are
assumed to be spherical with a fixed uniform size to avoid
complexity. Here we assume 1 $\mu$m-sized particles as found in the
HD142527 disk by \cite{honda09}, but the effect of the grain size is
discussed in \S4.3.4. The mass fractions of silicate and water to the
disk gas are assumed to be 0.0043 and 0.0094, respectively,
\citep{miyake93} and uniform throughout the disk. This water abundance
corresponds to the number fraction of H$_2$O molecules relative to
all gas particles in the disk as $X^{\rm H_2O}_{\rm total}=1.2\times10^{-3}$
which is the maximum fraction of water vapor.\footnote{The H$_2$O
abundance in \cite{miyake93} was not taken into account the recent
downward revision of O abundance \citep{asp09}.} 
The absorption and scattering coefficients of silicate and water ice
particles are also taken from \cite{miyake93}. The coefficients for 1
\micron-sized particles for a unit disk mass are shown in Figure~1.

\begin{figure}
% \epsscale{0.5}
 \plotone{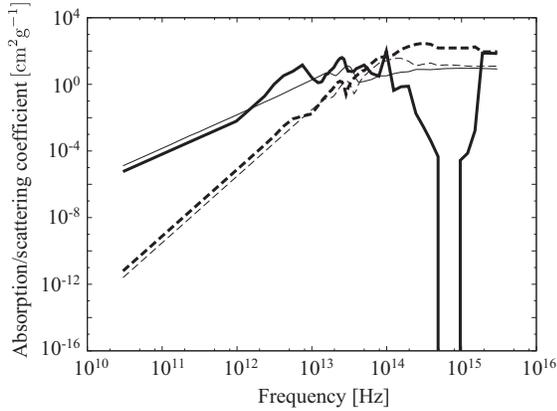}
 \caption{Absorption and scattering coefficients of the disk gas
 containing 1 $\mu$m-sized silicate and water ice particles adopted
 in our calculation. These coefficients are taken from Miyake \&
 Nakagawa (1993). Note that all the water molecules are contained in the
 icy dust particles. The thick and thin curves represent the
 coefficients by icy and silicate dust particles, respectively. The
 solid and dashed curves represent the absorption and scattering
 coefficients, respectively.\label{opacity_table_1micron}}
\end{figure}

\subsection{Ice abundance}

We evaluate the abundance of water ice in the disk by the balance
between changing rates of the ice dust radius due to condensation
$\dot{s}_{\mathrm{con}}$, sublimation $\dot{s}_{\mathrm{sub}}$, and
photodesorption $\dot{s}_{\mathrm{pd}}$;  
\begin{equation}
 \dot{s}_{\mathrm{sub}}(T) + \dot{s}_{\mathrm{con}}(T,X^{\rm H_2O}_{\rm{vapor}}) 
  + \dot{s}_{\mathrm{pd}} = 0, 
  \label{eq:balance_dust_size}
\end{equation}
where $X^{\rm H_2O}_{\rm{vapor}}$ is the number fraction of water vapor
of the disk medium and $T$ is the dust (and gas) temperature. The rates
are evaluated by the formulas from \cite{grigorieva07} (see Appendix
B). The dust temperature $T$ is obtained by the radiative transfer
calculation described in \S2.1 and Appendix A.

The number fraction of water vapor $X^{\rm H_2O}_{\rm{vapor}}$ is
determined as equation (\ref{eq:balance_dust_size}) is fulfilled. Then,
the number fraction of $\mathrm{H_{2}O}$ in water ice to the disk
gas $X^{\rm H_2O}_{\rm{ice}}$ is given by the residual of the
subtraction of $X^{\rm H_2O}_{\rm{vapor}}$ from the total number
fraction of H$_2$O $X^{\rm H_2O}_{\rm total}$. On the other hand, if 
$\dot{s}_{\rm{sub}} + \dot{s}_{\rm{con}} + \dot{s}_{\rm{pd}} < 0$ with 
$X^{\rm H_2O}_{\rm{vapor}} = X^{\rm H_2O}_{\rm total}$, water ice does
not condense and $X^{\rm H_2O}_{\rm{ice}} = 0$. Using 
$X^{\rm H_2O}_{\rm{total}}$ and $X^{\rm H_2O}_{\rm{ice}}$, the mass
ratio of water ice to the total water abundance $x_{\rm{ice}}$ is given
by  
\begin{equation}
 x_{\rm{ice}} = \frac{X^{\rm H_2O}_{\rm{ice}}}{X^{\rm H_2O}_{\rm total}}.
\end{equation}
Then, the absorption and scattering coefficients of a unit disk mass
containing dust particles, $\kappa_{\mathrm{abs}}$ and
$\kappa_{\mathrm{sca}}$ are given by  
\begin{eqnarray}
 \kappa_{\mathrm{abs}} &=& \kappa_{\mathrm{sil, abs}} 
  + x_{\mathrm{ice}} \kappa_{\mathrm{ice, abs}}, \\
 \kappa_{\mathrm{sca}} &=& \kappa_{\mathrm{sil, sca}} 
  + x_{\mathrm{ice}} \kappa_{\mathrm{ice, sca}},
\end{eqnarray}
where $\kappa_{\mathrm{sil,abs}}$, $\kappa_{\mathrm{sil,sca}}$,
$\kappa_{\mathrm{ice,abs}}$, and $\kappa_{\mathrm{ice,sca}}$ are the
absorption and scattering coefficients of silicate and water ice dust
particles by \cite{miyake93}, respectively ($\kappa_{\mathrm{ice,abs}}$
and $\kappa_{\mathrm{ice,sca}}$ are the coefficients in the case where
all the water molecules are condensed as icy dust particles). Since
$x_{\mathrm{ice}}$ itself affects the disk structure through the
radiative transfer, we solve $x_{\mathrm{ice}}$ and the disk structure
iteratively to obtain consistent solutions of them.

\subsection{Stellar and interstellar FUV radiation}

We consider three sources of UV radiation: the stellar photosphere,
an excess stellar radiation (probably caused by an accretion shock), and
the interstellar radiation field. The stellar photospheric radiation is
treated in the radiative transfer code described in \S2.1 and Appendix
A. We assume the spectrum to be a Planck function. The transfer of the
other two radiation sources are treated as described in the
subsequent paragraphs. As found in \S2.5 and \S3.2, however, the two
sources have a very weak effect on the position of the surface
snow line. The heating by the two sources is taken into account in
equation (A8) in addition to that of the photospheric radiation.

An excess FUV radiation is often observed in T Tauri stars, while
the photospheric radiation is dominant in Herbig stars
\citep[e.g.,][]{valenti03}. Nevertheless we consider the excess
radiation for discussions of the dependence of the stellar effective
temperature in \S3.2. We assume the excess FUV luminosity to be a
fraction of $10^{-3}$ to the bolometric luminosity of the photosphere
($L_*$) as observed in T Tauri stars \citep{herczeg04}. This reproduces
the observational trend that the excess relative to the photospheric
radiation is less important for earlier spectral type. Observations show
that the excess FUV radiation is dominated by Ly$\alpha$ photons
\citep{herczeg04}. We assume that the excess radiation consists only of
Ly$\alpha$ photons for simplicity. However, we omit any scattering of
the Ly$\alpha$ photons (see \S2.2) although the resonant
scattering of the photons is important in the sub-surface layer of
the disk \citep{fog11,bet11}. The Ly$\alpha$ assumption is only 
affects the calculation of the optical depth in the disk and the photon
number flux. Consequently, the excess FUV radiation flux in the disk can
be obtained as
\begin{equation}
 F_{\mathrm{FUV, excess}} (R, Z) = F_{\mathrm{FUV, excess, 0}} (R) 
  \exp(-\tau_{\mathrm{FUV}}(Z) / \beta), 
\end{equation}
where $F_{\mathrm{FUV,excess,0}} (R) = 10^{-3}L_{*} / (4\pi R^{2})$
is the incident excess FUV flux from the central star at the radius
$R$ for no dust case, $\tau_{\mathrm{FUV}}(Z)$ is the
absorption optical depth at the FUV ($\mathrm{Ly\alpha}$) wavelength
measured along the $Z$-axis from $(R, +\infty$) to $(R, Z)$, and
$\beta$ is the grazing angle for the stellar radiation.

We also consider the FUV radiation from the interstellar field. 
We adopt an average FUV radiation flux from \cite{habing68}.
The FUV flux is evaluated by assuming that it penetrates into the disk
parallel to the $Z$-axis: 
\begin{equation}
 F_{\mathrm{FUV,is}} (R, Z) = F_{\mathrm{FUV,is, 0}} 
  \exp(-\tau_{\mathrm{FUV}} (Z)), 
\end{equation}
where $F_{\mathrm{FUV,is, 0}}(R,Z)$ is the incident interstellar FUV
radiation flux and is set to be $1.6\times10^{-3}$ erg cm$^{-2}$
s$^{-1}$. The interstellar FUV radiation is also assumed to be consisted
only of $\mathrm{Ly\alpha}$ photons as the stellar excess FUV.
This is simply in order to treat the two additional FUV radiation
sources homogeneously, whereas the interstellar radiation may be
described by a Planck function or a power-law function made by a
combination of Planck functions with various temperatures.

\subsection{Examples of disk vertical structure}

\begin{figure*}
 \epsscale{0.9}
 \plotone{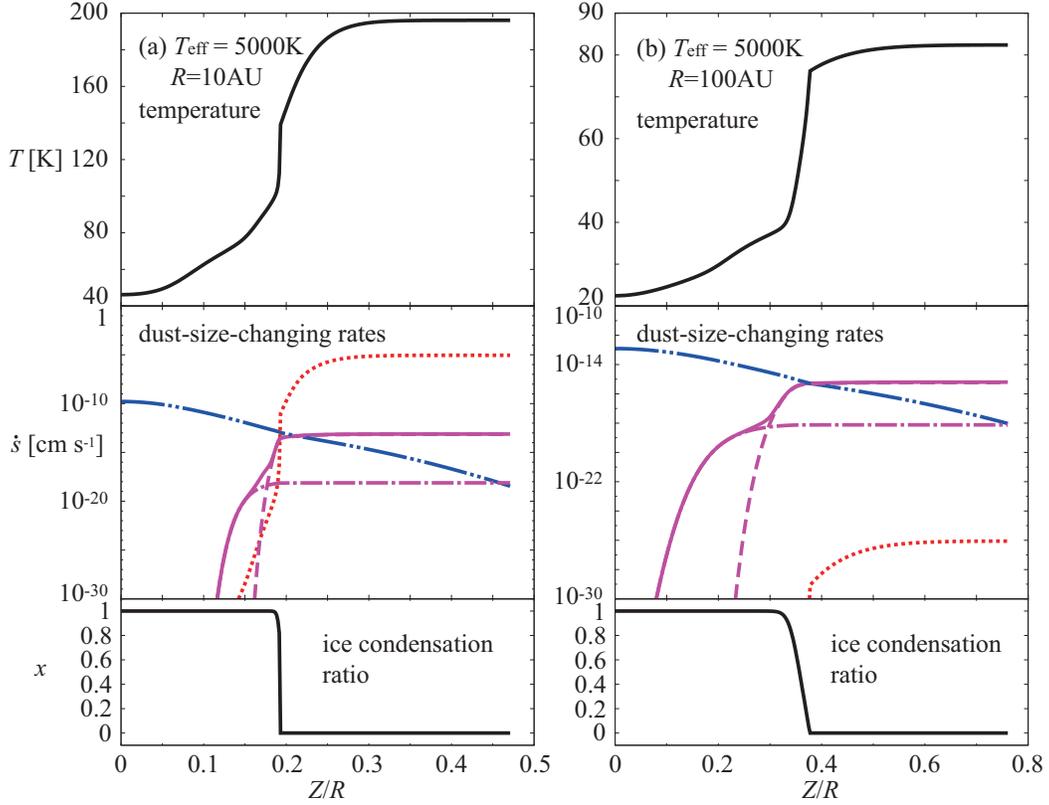}
 \caption{Vertical profiles of disk annuli with the radius of (a) 10 AU
 and (b) 100 AU. The stellar effective temperature of 5000 K, luminosity
 of 10 $L_\odot$, and mass of 1 $M_\odot$ are assumed. The top panels
 show the temperature profiles. The middle panels show the profiles of
 the changing rates of the grain size ${\dot s}$ due to condensation 
 (double-dot-dashed), sublimation (dotted), and photodesorption (solid). The
 dashed and dot-dashed curves represent the rates due to photodesorption
 induced by the stellar (photospheric and excess) FUV luminosity and the
 interstellar FUV radiation, respectively. The bottom panels show the
 profile of the ice condensation fraction $x$. The horizontal axis
 represents altitudes from the disk mid-plane scaled by the radius of the
 annuli.\label{struct}}
\end{figure*}

Figure \ref{struct} shows examples of the vertical structure of two
annuli at $R=10$ AU (left-hand tandem panels) and 100 AU 
(right-hand tandem panels) around a central star with the
effective temperature $T_*=5000$ K, the luminosity $L_*=10$ $L_\odot$,
and the mass $M_*=1$ $M_\odot$. The top panels show the temperature
structure along the disk vertical axis. The middle panels show the
changing rates of the grain size by condensation (double-dot-dashed
line), sublimation (dotted line), and photodesorption (solid line). We
also show the changing rates by photodesorption due to stellar
(photospheric and excess) radiation (dashed line) and due to
interstellar radiation (dot-dashed line). The bottom panels show the ice
condensation fraction.

In both annuli, condensation is rapid near the mid-plane and
virtually all water is in ice. As the temperature increases (and the
density decreases) along the vertical height, condensation becomes
slower and slower. On the other hand, sublimation and photodesorption
become suddenly effective beyond a certain height. Then, the ice
condensation front is formed at a height where condensation balances
with sublimation or photodesorption. In the $R=10$ AU case, sublimation
is faster than photodesorption at the front, and then, the balance is
established between condensation and sublimation. On the other hand, in
the $R=100$ AU case, sublimation is more than ten orders of magnitudes
slower than photodesorption because of a low temperature. Thus,
condensation balances with photodesorption.

Photodesorption by stellar FUV radiation always dominates that by
interstellar FUV radiation at the disk surface because of a much
stronger flux of the former radiation. However, the latter dominates the
former at a sub-surface layer. This is caused by the different incident
angles of the two radiations; the stellar radiation enters the disk with
a small angle against the surface (i.e. grazing angle $\beta$) and the
penetration depth becomes shallow, whereas the interstellar radiation
enters the disk vertically and reaches a deeper layer. When discussing
the surface snow line, we conclude that the interstellar radiation
hardly contributes to the determination of the surface snow
line. Therefore, we omit the contribution by the interstellar FUV
radiation in the remaining part of this paper.

\subsection{Definition of the surface snow line}

Let us define the position of the surface snow line, here. 
We consider the method observing the surface snow line by using the
3 $\mu$m H$_2$O ice absorption feature in the scattered light proposed
by \cite{inoue08} theoretically and attempted by \cite{honda09}
observationally. The 3 $\mu$m scattered light comes from the layer to
which the optical depth for the 3 \micron\ radiation from the
central star is about unity. Therefore, we define the 
layer as the disk `surface' or `3 $\mu$m surface' in this paper. Next,
we consider the ice-condensation front in the $(R,Z)$-plane. Without the
accretion viscous heating, the temperature decreases along the radial
coordinate $R$ and increases along the vertical coordinate $Z$. Thus,
the ice-condensation front becomes lower $Z$ for smaller $R$ and higher
$Z$ for larger $R$ (see Fig.~\ref{2d_snowline}). Then, there may be a
cross-point between the 3 $\mu$m surface and the ice-condensation
front. If we observe a disk with the scattered light, we will observe
the 3 $\mu$m feature only from the outside of the cross-point. 
Therefore, we define the cross-point as the surface snow line in this
paper. Namely, the surface snow line in this paper is the snow line
which would be observed by the 3 $\mu$m scattered light.

\section{Result}

\subsection{Structure of the ice-condensation front and the surface snow line}

\begin{figure*}
 \epsscale{0.9}
 \plotone{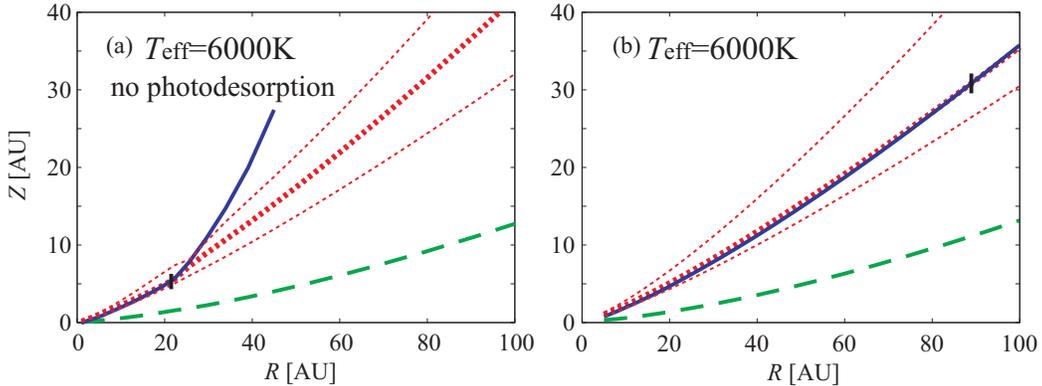}
 \caption{The structure of the ice-condensation front in a circumstellar
 disk. The horizontal and vertical axes represent the distance from the
 central star $R$ and the height from the mid-plane $Z$ in astronomical
 unit, respectively. The panels (a) and (b) are the cases without and
 with photodesorption, respectively. The stellar parameters assumed
 are the effective temperature of 6000 K, the luminosity of 10 $L_\odot$
 and the mass of 1 $M_\odot$. The dust particle size is assumed to be
 1.0 $\micron$. The solid curve represents the condensation front below
 which water ice particles exist. The dashed curve represents the
 pressure scale height of the disk medium. The upper (thin), middle
 (thick), and lower (thin) dotted curves represent layers at which the
 optical depth at wavelength 3 $\mu$m from the central star becomes 0.1,
 1 (3 \micron\ `surface'), and 10, respectively. The vertical tick
 indicates the cross-point between the ice-condensation front and the
 3-$\mu$m disk surface, i.e. the surface snow line position. 
 \label{2d_snowline}}
\end{figure*}

Figure \ref{2d_snowline} shows the ice-condensation front (solid
curve) and the surface snow line position (vertical tick) in a disk
around a star with the effective temperature $T_*=6000$ K, the
luminosity $L_*=10$ $L_{\odot}$, and the mass $M_*=1$ $M_\odot$.
The layers of $\tau_{3\mu{\rm m}}=0.1$, 1, and 10 are also shown as the
dotted curves, where $\tau_{3\mu{\rm m}}$ is the optical depth at the 
wavelength 3 \micron\ from the central star. The layer of 
$\tau_{3\mu{\rm m}}=1$ is the 3 \micron\ surface of the disk. We also
show the pressure scale-height calculated by the mid-plane temperature
as the dashed curve in each panel.

Comparing the panel (a), where photodesorption is omitted,
with the panel (b), where the process works, we find that
photodesorption wipes out water ice particles above the disk surface and
the ice-condensation front almost coincides with the surface throughout
the disk. The position of the surface snow line (i.e. the cross-point of
the ice-condensation front and the 3 \micron\ surface) shifts outward by
a factor of about 4 for this case. As shown later (Figs.
\ref{2d_snowline_ice_vs_noice} and \ref{snowline_vs_Teff_M1}), the
radial shift becomes larger and larger as the effective temperature
increases. Therefore, photodesorption is very important to determine the
position of the surface snow line for Herbig stars.

\subsection{Stellar parameter dependence of the surface snow line}

\begin{figure*}
 \epsscale{0.9}
 \plotone{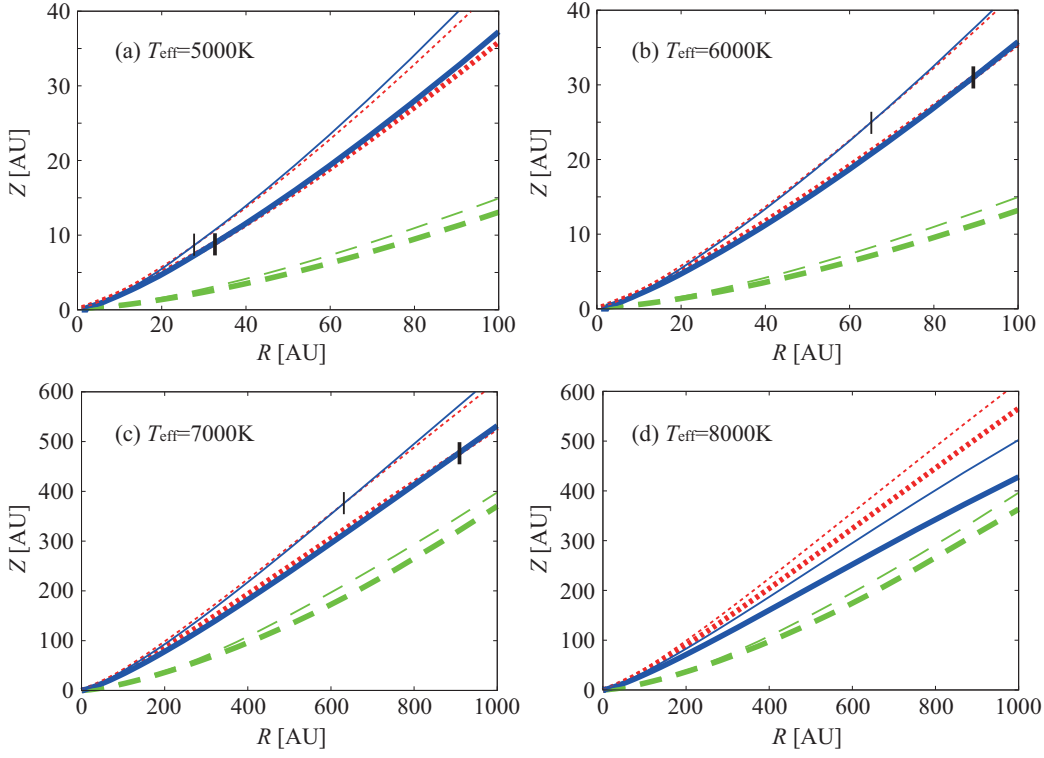}
 \caption{Same as Figure \ref{2d_snowline}, but for various stellar
 effective temperatures. The thick and thin curves are the cases with
 and without the water ice opacity, respectively. Note that there is a
 difference in the axis scales between the panels (a) and (b) and the
 panels (c) and (d). \label{2d_snowline_ice_vs_noice}}
\end{figure*}

\begin{figure}
 % \epsscale{0.9}
 \plotone{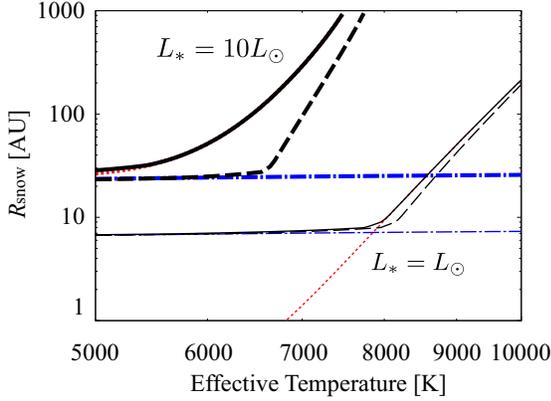}
 \caption{Position of the surface snow line as a function of stellar
 effective temperature with the stellar mass of 1 $M_{\odot}$. The
 horizontal and vertical axes represent the effective temperature and
 the position of the surface snow line, respectively. The thick and thin
 curves show different stellar luminosities as indicated in the
 panel. The dot-dashed curves show the snow line position where the
 equilibrium between sublimation and condensation is established.
 The dotted curves show the position where the equilibrium between
 photodesorption and condensation is established. The solid curves show
 the position determined by the equilibrium among photodesorption,
 condensation, and sublimation. The dashed curves are the same as the
 solid curves but without the stellar UV excess. \label{snowline_vs_Teff_M1}}
\end{figure}

In this section, we examine the dependence of the surface snow line 
position $R_{\rm snow}$ on the stellar effective temperature $T_*$, the
luminosity $L_*$, and the mass $M_*$. In this calculation, we omit
opacity of water ice particles for the computational cost. This
simplification gives 20--30\% smaller radius of the surface snow line
position as shown in Figure \ref{2d_snowline_ice_vs_noice}. 
This relatively small difference is probably caused by an offset of
two opposite effects; (1) A smaller opacity in the case without ice
opacity results in a higher temperature in the disk interior (i.e. a
higher pressure height found in Fig.~\ref{2d_snowline_ice_vs_noice}) 
and also a higher surface height defined by $\tau_{\rm 3\mu m}=1$ 
(as also found in Fig.~\ref{2d_snowline_ice_vs_noice}). However, this
leads a smaller density at the surface, and then, the condensation rate
becomes slower and the surface snow line shifts outward. (2) The smaller
opacity also results in a larger column density from infinity to the
surface in order to account for $\tau_{\rm 3\mu m}=1$ at the surface. It
leads a lower surface height and a larger density at the surface. Then, 
the condensation rate becomes faster, and the surface snow line shifts
inward. This second effect exceeds the first effect in the cases shown
in Figure~\ref{2d_snowline_ice_vs_noice}. In any case, the omission of
the ice opacity does not change the mechanism determining the surface
snow line position.


\begin{figure}
 % \epsscale{0.9}
 \plotone{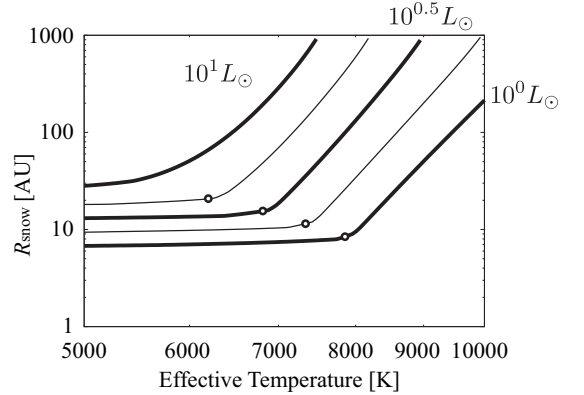}
 \caption{Same as Figure \ref{snowline_vs_Teff_M1}, but for various
 stellar luminosities. The open circles indicate
 the critical temperatures $T_{\rm c}$ defined in \S3.2.
 \label{snowline_vs_Teff_with_Luminosy}}
\end{figure}

\begin{figure}
 % \epsscale{0.9}
 \plotone{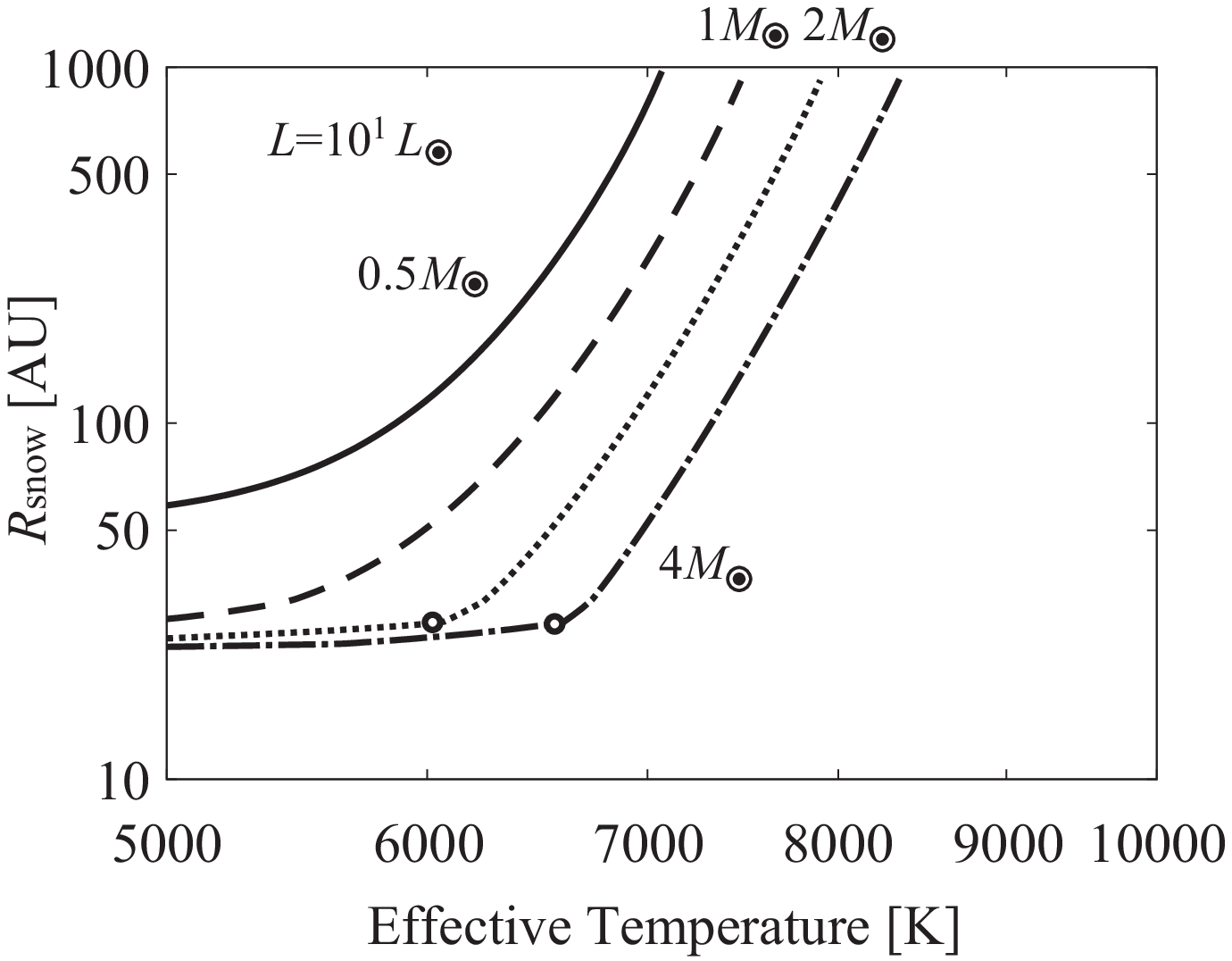}
 \caption{Same as Figure \ref{snowline_vs_Teff_M1}, but for various
 stellar masses. The stellar luminosity of 10 $L_{\odot}$ is assumed. The
 solid, dashed, dotted, and dot-dashed curves show the surface snow line
 position with the stellar mass of 0.5 $M_{\odot}$, 1 $M_{\odot}$,
 2 $M_{\odot}$, and 4 $M_{\odot}$, respectively. The open circles indicate
 the critical temperatures $T_{\rm c}$ defined in \S3.2.
 \label{snowline_various_mass}}
\end{figure}

\subsubsection{Effective temperature}

As found in Figure \ref{2d_snowline_ice_vs_noice}, the position of
the surface snow line strongly depends on the stellar effective
temperature; as the temperature increases, the surface snow line shifts
outward greatly and eventually disappears. Figure
\ref{snowline_vs_Teff_M1} shows this dependence more in detail for the
cases with $M_*=1$ $M_{\odot}$ and $L_*=1$ $L_\odot$ and 10 $L_\odot$.
The dot-dashed curves indicate the cases without photodesorption,
i.e., $R_{\rm snow}$ is determined by the balance between sublimation
and condensation. On the other hand, the dotted curves are the
cases without sublimation, i.e., $R_{\rm snow}$ is determined by the
balance between photodesorption and condensation. The solid curves
are the cases with all the three processes. The dashed curves are
same as the solid curves but without the stellar excess FUV (see \S2.4).

For the sublimation--condensation cases (dot-dashed curves), 
$R_{\rm snow}$ is about 7 or 25 AU and almost independent of $T_*$, but
depends on $L_*$ as $R_{\rm snow}\propto L_*^{1/2}$. This is because the
gas temperature for the sublimation--condensation balance is
almost constant at $\sim100$ K at the disk surface where the dust
temperature is determined only by bolometric flux (i.e. $\propto L_*/R^2$) 
and not the spectrum (see also \S4.2). For the
photodesorption--condensation cases (dotted curves), $R_{\rm snow}$
strongly depends on $T_*$ as $R_{\rm snow} \propto T_*^{12}$. This
strong dependence comes from the $T_*$ dependence in the FUV photon
number flux (see \S4.2). It also depends on $L_*$ as 
$R_{\rm snow}\propto L_*^{2}$ which is different from the
sublimation--condensation case.

When all the three processes are taken into account (solid
curves), $R_{\rm snow}$ traces the sublimation--condensation
case for lower $T_*$ and the photodesorption--condensation case for
higher $T_*$. In other words, the mechanism determining $R_{\rm snow}$ 
is the sublimation--condensation balance for lower $T_*$ and the
photodesorption--condensation balance for higher $T_*$. We here define a
critical temperature $T_{\rm c}$ as $T_*$ at the intersection point
between the two curves of the sublimation--condensation case and the
photodesorption--condensation case. If $T_*>T_{\rm c}$, photodesorption
is efficient and the surface snow line moves to the outermost part of
the disk or disappears. Thus, $T_{\rm c}$ is an indicator for judging
the possibility to detect the 3 \micron\ water ice absorption feature in
the disk scattered radiation (see \S4.1).

If we compare the dashed curves (the cases without the excess FUV) with
the solid curves, we can see the effect of the stellar FUV excess on the
surface snow line. Although it is still uncertain how strong the FUV
excess is, we assume that the FUV excess accounts for only 0.1\% of
$L_*$ (\S2.4). We find an effect of the FUV excess in the $L_*=10$
$L_\odot$ case. If the FUV excess is stronger than this level, we
should take it into account in order to determine the position of the
surface snow line correctly.

\subsubsection{Stellar luminosity}

Figure \ref{snowline_vs_Teff_with_Luminosy} shows $R_{\rm snow}$ as a
function of $T_*$ with various $L_*$s and $M_*=1$ $M_{\odot}$. It is
seen that $R_{\rm snow}$ shifts outward as $L_*$ increases. This is
because both the FUV radiation flux from the central star and the
temperature in the disk increase as $L_*$ increases. It is also seen
that the critical temperature $T_{\rm c}$ becomes lower as $L_*$
increases. The dependence is approximately 
$T_{\rm c} \propto L_*^{-1/8}$ (see \S4.2).

\subsubsection{Stellar mass}

Figure \ref{snowline_various_mass} shows $R_{\rm snow}$ as a function of
$T_*$ with $L_*=10$ $L_{\odot}$ and various $M_*$s. It is seen that the
critical temperature $T_{\rm c}$ becomes higher as $M_*$ increases. This
is because the condensation rate is approximately proportional to the
square root of $M_*$; the condensation rate depends linearly on the gas
pressure with the assumption of uniformly constant number fraction
of water molecules in the whole disk, and the gas pressure depends
approximately linearly on the square root of $M_*$ 
(see eqs.~A1--A3, and also \S4.2).

\section{Discussion}

\subsection{Implications for the observation}

In this section, we discuss the effect of photodesorption on observing
the surface snow line. As proposed by \cite{inoue08}, we can
resolve the surface snow line by using the 3 \micron\ water ice
absorption feature in the scattered light if we have a good enough
spatial resolution. This method implicitly assumes the presence of water
ice particles at the 3 \micron\ disk surface. However, an efficient
photodesorption may remove the particles from the surface and the
absorption feature from the scattered radiation. On the other hand,
\cite{honda09} detected the absorption feature in the scattered light
from the outer disk at $R\ga140$ AU of HD142527, although they did not
resolve the snow line position because of insufficient spatial
resolution. Therefore, photodesorption does not remove water ice
particles completely from the disk surface of HD142527.

\begin{figure}
 % \epsscale{0.9}
 \plotone{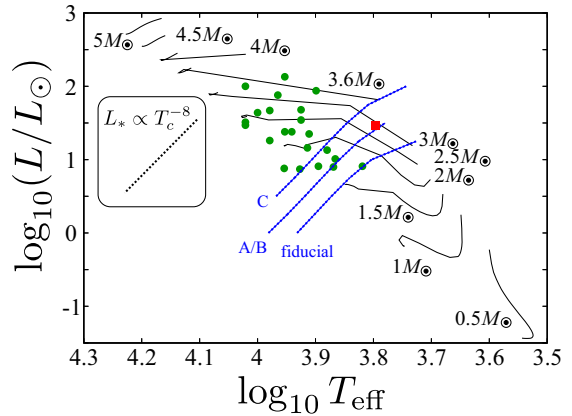}
 \caption{Boundaries (dotted lines with fiducial, A/B, and C)
 at which the mechanism determining the position of the surface snow
 line changes. In the upper left side from the boundaries, the mechanism
 is the balance between photodesorption and condensation. In the lower
 right side, the mechanism is the balance between sublimation and
 condensation. The labels fiducial, A/B, and C correspond to the models
 summarized in Table \ref{table:parameter}. Note that the models A and B
 are completely overlapped. The inset shows the slope of the boundaries
 analytically derived in \S4.2. The solid curves with labels of the
 stellar mass show the evolution tracks of pre-main sequence stars from
 Yi et al.~(2001). The circles are the stellar parameters of the Herbig
 Ae/Be stars from van Boekel et al.~(2005). The square shows HD142527
 observed by Honda et al.~(2009). \label{HR1}}
\end{figure}

As found in \S3.2.1, photodesorption surpasses sublimation at the
disk surface and $R_{\rm snow}$ is determined by the balance between
photodesorption and condensation when $T_*$ exceeds a certain value
which is defined as the critical temperature $T_{\rm c}$. This
temperature also depends on $L_*$ as found in \S3.2.2. Let us over-plot
the boundaries at which the mechanism determining $R_{\rm snow}$ 
changes from the sublimation--condensation balance to the
photodesorption--condensation balance on the HR diagram and compare
them with Herbig Ae/Be stars on the diagram (Fig.~\ref{HR1}). 
In particular, HD142527, where the water ice feature in the
scattered light was detected, is the key object in this
comparison. In addition, we also examine effects of other parameters
such as photodesorption yield, H$_2$O abundance, etc.\ on the
boundaries.

As a reference, we define the fiducial model where $M_*=3$ $M_{\odot}$,
the surface density at 1 AU $\Sigma_{1\rm{AU}}=1700~\rm{g\ cm^{-2}}$,
the total (vapor+ice) number fraction of water 
$X_{\rm total}^{\rm H_2O} = 1.2\times 10^{-3}$, and the photodesorption
yield $Y_{\mathrm{pd}} = 1\times10^{-3}$. In addition, we consider
three other models: A, B, and C, where $X_{\rm total}^{\rm H_2O}$ is
increased by a factor of 2, $Y_{\rm pd}$ is reduced by a factor of 2,
and $X_{\rm total}^{\rm H_2O}$ is increased by a factor of 2 and 
$Y_{\rm pd}$ is reduced by a factor of 2, respectively (Table~1).
In Figure~\ref{HR1}, we show the boundaries of these models by the
dotted lines with the labels like `fiducial' on the HR diagram.
Interestingly, the boundaries of the models
A and B are completely overlapped. The reason is the offset of twice 
$X_{\rm total}^{\rm H_2O}$ in the model A and half $Y_{\rm pd}$ in the
model B as shown in \S4.2 analytically. In Figure \ref{HR1}, we also
show the evolution tracks of pre-main sequence stars with various
stellar masses from \cite{yi01} (solid lines with labels of $M_*$) 
and the stellar parameters of the observed Herbig Ae/Be
stars from \cite{van_voekle05} (circles). HD142527 is indicated by
the square.

In the upper-left region from the boundaries in Figure~\ref{HR1},
the photodesorption--condensation balance determines $R_{\rm snow}$. 
On the other hand, in the lower-right region from the boundaries, the
sublimation--condensation balance determines it. In the
photodesorption--condensation case, the surface snow line moves outward
significantly as shown in \S3.2. Since \cite{honda09} detected ice
particles on the disk surface at $R\gtrsim140$ AU of HD142527, we expect
that HD142527 is in the sublimation--condensation area. However, it is
in the photodesorption--condensation area for the fiducial model. On the
other hand, this object is in the sublimation--condensation area for the
model C and marginally for the models A and B.

\begin{figure}
 % \epsscale{0.9}
 \plotone{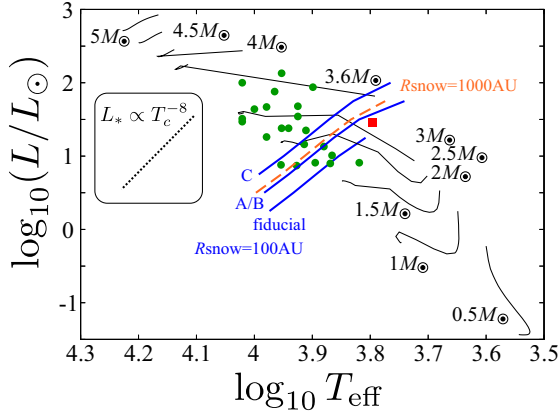}
 \caption{Same as Figure \ref{HR1}, but the solid lines with
 fiducial, A/B, and C are the sets of the stellar luminosity and
 effective temperature which give the position of the surface snow line
 $R_{\rm snow}=100$ AU. The dashed line is the sets which give 
 $R_{\rm snow}=1000$ AU for the fiducial model. \label{HR2}}
\end{figure}

Even if the surface snow line is controlled by photodesorption, it
is possible that $R_{\rm snow}<140$ AU as shown in
Figures~\ref{snowline_vs_Teff_M1} and \ref{snowline_vs_Teff_with_Luminosy}.
Then, we show the boundaries at which $R_{\rm snow}=100$ AU on the HR
diagram in Figure~\ref{HR2}. The solid lines with labels fiducial, A/B,
and C are the boundaries for the four models in
Table~\ref{table:parameter}. We also show the boundary of 1000 AU for
the fiducial model as the dashed line for a comparison. We find HD142527
between the boundaries of 100 AU for the fiducial and A/B models. In
fact, this object is on the boundary of 300 AU for the fiducial model
which is not shown in Figure~\ref{HR2}. Given the uncertainty of our
simple model, we conclude that $R_{\rm snow}$ in the HD142527 disk is
100--300 AU which is still consistent with the observations by
\cite{honda09}.

On the other hand, there are many Herbig Ae/Be stars in the upper-left
region far from the boundaries of 100 AU and even 1000 AU 
in Figure~\ref{HR2}. Photodesorption is probably very efficient for
these objects, and water ice particles disappear from the surface of the
disk around the objects. Then, we expect that there is not the 3
\micron\ water ice feature in the scattered light from the disks. This
will be an observational test to examine whether photodesorption is
active or not on the disk surface.

Finally we note a possible caveat in resolving the surface snow line by
using the water ice 3 \micron\ feature in the scattered light. As found
in Figure \ref{2d_snowline_ice_vs_noice}, the ice-condensation front and
the 3 \micron\ surface (i.e. $\tau_{\rm 3\mu m}=1$ layer) are always
close to each other when $T_*\la T_{\rm c}$. When we are moving from the
outer disk to the inner one on the scattered light image, thus, the
strength of the 3 \micron\ feature may not disappear suddenly at the
surface snow line, but may weaken gradually. In other words, the surface
snow line is not a `clear' line. In order to predict the change of the
feature strength, we need to simulate observations, which would be a
future work.

\begin{table*}
\begin{center}
\caption{Parameters for the $(T_{\mathrm{c}}, L_{*})$ curves in Figures
 \ref{HR1} and \ref{HR2}.\label{table:parameter}}
\begin{tabular}{ccccc}
\tableline\tableline
Model & $\Sigma_{\mathrm{1AU}}$ (g cm$^{-2}$) & $M_{*}$ ($M_\odot$) 
 & $X^{\rm H_2O}_{\rm total}$ & $Y_{\mathrm{pd}}$ \\
\tableline 
fiducial model & 1700 & 3 & $1.2\times 10^{-3}$ & $1.0\times 10^{-3}$ \\
A & 1700 & 3 & $2.4\times 10^{-3}$  & $1.0 \times 10^{-3}$ \\
B & 1700 & 3 & $1.2\times 10^{-3}$  & $0.5 \times 10^{-3}$ \\
C & 1700 & 3 & $2.4\times 10^{-3}$  & $0.5 \times 10^{-3}$ \\
\tableline
\end{tabular}
%% Any table notes must follow the \end{tabular} command.
%\tablenotetext{a}{Sample footnote for table~\ref{tbl-2} that was
%generated with the \LaTeX\ table environment}
%\tablenotetext{b}{Yet another sample footnote for table~\ref{tbl-2}}
%\tablenotetext{c}{Another sample footnote for table~\ref{tbl-2}}
%\tablecomments{We can also attach a long-ish paragraph of explanatory
%material to a table.}
\end{center}
\end{table*}

\subsection{Analytic expression of the critical effective temperature}

Here we try to derive an analytic formula describing the critical
temperature, $T_{\rm c}$. The mechanism determining $R_{\rm snow}$
changes depending on $T_*$; if $T_*<T_{\rm c}$, it is the balance
between condensation and sublimation, whereas if $T_*>T_{\rm c}$, it is
the balance between condensation and photodesorption. Thus, equation 
(\ref{eq:balance_dust_size}) can be approximated to 
\begin{equation}
 \cases{
  \dot{s}_{\rm con}+\dot{s}_{\rm sub}\approx 0 & ($T_*<T_{\rm c}$) \cr
  \dot{s}_{\rm con}+\dot{s}_{\rm pd}\approx 0 & ($T_*>T_{\rm c}$) \cr 
  }\,.
  \label{eq:sizechangerate2}
\end{equation}

When $T_*<T_{\rm c}$, equation (\ref{eq:sizechangerate2}) becomes 
$X^{\rm H_2O}_{\rm total} P \approx P_{\rm sat}$, where 
$X^{\rm H_2O}_{\rm total}$ is the number fraction of H$_2$O molecules
relative to all gas particles, $P$ and $P_{\rm sat}$ are the gas 
pressure and the saturation pressure of water vapor, respectively  
(see eqs.~[B1--B5]). We have assumed that the vapor fraction 
$X^{\rm H_2O}_{\rm vapor}\sim X^{\rm H_2O}_{\rm total}$. The saturation
pressure is a rapid function of gas temperature $T$. Thus, 
$X^{\rm H_2O}_{\rm total} P \approx P_{\rm sat}$ is established only
when $T\simeq100$ K on the disk surface. This means that $R_{\rm snow}$
is the position where the dust temperature becomes $T\simeq100$ K. The
dust temperature on the disk surface is determined by 
$\sigma T^4 \approx L_*/4\pi R^2$. For a fixed $T$, we obtain 
$R_{\rm snow}$ determined by the condensation--sublimation balance as
\begin{equation}
 R_{\rm snow,sub} \propto L_*^{1/2}\,.
\end{equation}
This relation is very consistent with the dot-dashed lines in 
Figure \ref{snowline_vs_Teff_M1}.

When $T_*>T_{\rm c}$, equation (\ref{eq:sizechangerate2}) becomes 
$Y_{\rm pd} N_{\rm abs} \propto X^{\rm H_2O}_{\rm total} \rho$, 
where $Y_{\rm pd}$ is the photodesorption yield, $N_{\rm abs}$ is
the number flux of FUV photons absorbed by water ice grains, and $\rho$
is the gas mass density (see eqs.~[B4--B6]). The FUV number flux
$N_{\rm abs}$ given by equation (B7) can be approximated to 
$N_{\rm abs}\propto L_* T_*^6 R^{-2}$ for a fixed $L_*$ when the FUV
absorption efficiency parameter (ratio of the absorption and
geometrical cross sections of grains) $Q_{\rm abs}\sim1$ and the
stellar FUV excess is omitted. The gas density $\rho$ should be that at
the 3 \micron\ surface:  
$\rho_{\rm 3\mu m}=\rho(\tau_{\rm 3 \mu m}=1)$. This can be approximated
to $\rho_{\rm 3\mu m}\propto 1/H_{\rm P} \propto M_*^{1/2}R^{-3/2}$, 
where $H_{\rm P}$ is the pressure scale-height. We have omitted the $R$
dependence in the equatorial temperature $T_0$ because $T_0$ is almost
constant at $R\sim100$ AU \citep{cg97}. Note that $\rho_{\rm 3\mu m}$ is
independent of the normalization of the surface mass density $\Sigma$
because we always see the layer with $\tau_{\rm 3\mu m}=1$. However, it
depends on $H_{\rm P}$ which determines the density profile. As a
result, we obtain 
$Y_{\rm pd} L_* T_*^6 R^{-2} \propto X^{\rm H_2O}_{\rm total} M_*^{1/2} R^{-3/2}$,
and then, $R_{\rm snow}$ determined by the condensation--photodesorption
balance becomes
\begin{equation}
 R_{\rm snow,pd} \propto T_*^{12} L_*^2 Y_{\rm pd}^2 
  {X^{\rm H_2O}_{\rm total}}^{-2} M_*^{-1}\,,
\end{equation}
which agrees with the dotted lines in Figure \ref{snowline_vs_Teff_M1}.

Finally, we define $T_{\rm c}$ as $T_*$ when 
$R_{\rm snow,sub}=R_{\rm snow,pd}$. Then we obtain 
\begin{equation}
 T_{\rm c} \propto L_*^{-1/8} M_*^{1/12} Y_{\rm pd}^{-1/6} 
  {X^{\rm H_2O}_{\rm total}}^{1/6}\,.
\end{equation}
Comparing this with the results shown in Figure \ref{HR1}, we obtain
\begin{eqnarray}
 T_{\rm c}=8500~{\rm K} 
  \left(\frac{L_*}{1~L_\odot}\right)^{-1/8} 
  \left(\frac{M_*}{3~M_\odot}\right)^{1/12} \cr \times
  \left(\frac{Y_{\rm pd}}{1\times10^{-3}}\right)^{-1/6}
  \left(\frac{X^{\rm H_2O}_{\rm total}}{1.2\times10^{-3}}\right)^{1/6}\,.
\end{eqnarray}
This formula would be useful for the readers to expect which balance
determines $R_{\rm snow}$. If $T_*>T_{\rm c}$, one may expect the balance
between condensation and photodesorption. For the opposite case,
the balance between condensation and sublimation is expected.

\subsection{Effects of processes omitted}

\subsubsection{Photodissociation}

We have omitted photodissociation of H$_2$O molecules, whereas this
process is suggested to be important in disks around Herbig stars
\citep{pon10,fed11}. Here we justify this omission.

The photodissociation rate is given by 
$k_{\rm pdis} \approx \sigma_{\rm pdis} N_{\rm FUV}$,
where $\sigma_{\rm pdis}$ is the cross section for photodissociation and
$N_{\rm FUV}$ is the number flux of FUV photons.\footnote{$N_{\rm abs}$
in the previous subsection is equal to $Q_{\rm abs}N_{\rm FUV}$.}
According to \cite{lee84}, $\sigma_{\rm pdis}\sim10^{-18}$ cm$^2$ for
H$_2$O molecules. On the other hand, the photodesorption rate is given
by $k_{\rm pdes} \approx \pi s^2 Q_{\rm abs} Y_{\rm pd} N_{\rm FUV}$,
where $s$ is the grain size, $Q_{\rm abs}\sim1$ is the absorption
efficiency parameter, and $Y_{\rm pd}$ is the photodesorption
yield. Since the time-scale is $\tau=1/k$, the ratio of the two
time-scales is 
$\tau_{\rm pdis}/\tau_{\rm pdes} = k_{\rm pdes}/k_{\rm pdis}
\approx\pi s^2 Y_{\rm pd}/\sigma_{\rm pdis} \sim 10^7$ for 
$s\sim1$ \micron\ and $Y_{\rm pd}\sim10^{-3}$. Then we find that 
photodissociation is several orders of magnitude slower than
photodesorption. Therefore, photodissociation does not affect the
balance between photodesorption and condensation.

We may still argue the importance of photodissociation in observations.
When photodesorption becomes faster than condensation, water ice
particles are finally evaporated. This destruction time-scale is given
by $\tau_{\rm dest}=s/|{\dot s}_{\rm pd}|$, where ${\dot s}_{\rm pd}$ is
the size changing rate by photodesorption (eq.~B6). The ratio of this
time-scale to the photodissociation one is  
$\tau_{\rm dest}/\tau_{\rm pdis}\approx\sigma_{\rm pdis}\rho_{\rm ice} s 
/(m_{\rm H_2O}Y_{\rm pd})\sim10^5$ for $s\sim1$ \micron\ and 
$Y_{\rm pd}\sim10^{-3}$ ($\rho_{\rm ice}$ is the ice material density and
$m_{\rm H_2O}$ is the mass of a H$_2$O molecule). Thus, when the ice
destruction is completed, evaporated H$_2$O molecules have been already
photodissociated. Therefore, we expect that there is no H$_2$O molecules
in gas phase above the disk surface where photodesorption destroys ice
particles.

\subsubsection{X-ray photodesorption}

We have neglected desorption by X-ray \citep[e.g.,][]{naj01}.
The photodesorption rates by X-ray and FUV are given by 
$k_i \approx \pi s^2 Q_i Y_i N_i$, where $i$ is X-ray or FUV, $s$
is the grain size, $Q_i$ is the absorption efficiency, $Y_i$ is the
photodesorption yield, and $N_i$ is the photon number flux. Thus, the
ratio is $k_{\rm X}/k_{\rm FUV} \approx (Q_{\rm X}/Q_{\rm FUV}) 
(Y_{\rm X}/Y_{\rm FUV})(N_{\rm X}/N_{\rm FUV})$. 

X-ray luminosity of Herbig stars is observed in a range of 
$10^{-7}$--$10^{-4}$ of the bolometric luminosity \citep{ste05}.
Given about two orders of magnitude higher energy per a photon in X-ray
than in FUV and that $\sim10$\% of the bolometric energy is emitted in
FUV for Herbig stars, we obtain $N_{\rm X}/N_{\rm FUV}<10^{-5}$. 
The X-ray absorption efficiency is $Q_{\rm X}\sim10^{-1}$ which is 
estimated from the sum of atomic photoionization cross sections. Then,
we find $Q_{\rm X}/Q_{\rm FUV}\sim10^{-1}$. The yield for X-ray 
$Y_{\rm X}$ is uncertain for water ice, while $Y_{\rm FUV}\sim10^{-3}$.
\cite{naj01} estimated $Y_{\rm X}\sim10$--$10^3$ for CO from a
consideration of `spot heating' of subunits which is hit by a X-ray
photon and is poorly connected thermally with other parts in a grain.
They considered $<0.01$ \micron\ subgrains, which cover the parent grain,
as the subunits. Comparing CO to H$_2$O, the latter has 5 times larger
binding energy than the former \citep{san88}, which reduces $Y_{\rm X}$
for H$_2$O by the same factor. In addition, the connecting `neck'
between H$_2$O ice particles is thicker than that of silica considered
in \cite{naj01} because of sintering \citep{sir11}, further reducing
$Y_{\rm X}$ because of an enhanced conductivity of heat. Thus, we
estimate $Y_{\rm X}/Y_{\rm FUV}<10^5$, although a more detailed study is
encouraged. In summary, we obtain $k_{\rm X}/k_{\rm FUV}<10^{-1}$ and we
can omit X-ray photodesorption at the disk surface. However, we
note that X-ray photodesorption should be important in the disk
interior, where $k_X/k_{\rm FUV}>1$ because $N_{\rm FUV}$ is
significantly smaller than that at the surface.

\subsubsection{Dust settling and turbulent mixing}

In this paper, we have not considered dust settling and turbulent mixing
although they may affect the vertical distribution of water ice
particles and the surface snow line. If the dust settling toward the
mid-plane occurs, the height of the 3 \micron\ surface is reduced. This
results in a larger density and a faster condensation at the
surface. Thus, the surface snow line moves inward. To estimate the
amount of the shift, we need further quantitative studies.

The turbulent mixing may supply water ice particles from the disk
interior to the disk surface even if photodesorption removes the
particles from the surface. However, the mixing time-scale is
$1\times10^3$ yr at 10 AU and $1\times10^4$ yr at 100 AU according to
\cite{hei11}, while the destruction time-scale of 1 \micron\ ice
particles by photodesorption is $3\times10^2$ yr at 10 AU and
$3\times10^3$ yr at 100 AU. Therefore, the ice supply by turbulence is
not fast enough.

\subsubsection{Grain size dependence}

\begin{figure}
 % \epsscale{0.5}
 \plotone{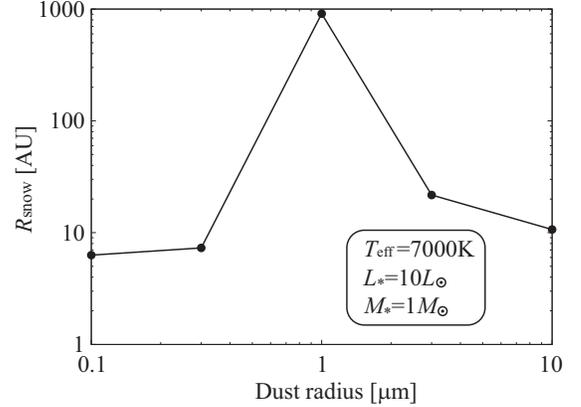}
 \caption{Position of the surface snow line as a function of dust
 grain size for the case of the effective temperature of 7000 K, the
 luminosity of 10 $L_\odot$, and the mass of 1 $M_\odot$.
 \label{grainsize}} 
\end{figure}

\begin{figure}
 % \epsscale{0.9}
 \plotone{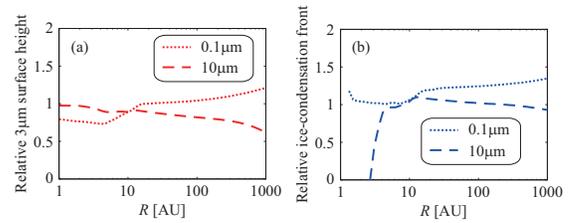}
 \caption{(a) Height of the 3 \micron\ surface relative to the 1
 \micron-sized grain case. (b) Height of the ice-condensation front
 relative to the 1 \micron-sized grain case. The dotted and dashed lines
 are the cases of 0.1 \micron-sized and 10 \micron-sized grain cases,
 respectively. \label{snowline_grainsize}}
\end{figure}

Figure \ref{grainsize} shows $R_{\rm snow}$ as a function of dust
grain size. We find that $R_{\rm snow}$ depends on the grain size {\bf
in a complex way}; the 1 \micron-sized grain case gives the largest 
$R_{\rm snow}$ and smaller or larger grain size cases give smaller 
$R_{\rm snow}$s. This is because $R_{\rm snow}$ is the cross-point of
the two curves, i.e. the ice-condensation front and the 3 \micron\ disk
surface, which have a similar slope, and it is very sensitive to the
relative position of the two curves. 

To see the reason of this complex behavior of $R_{\rm snow}$, we show in 
Figure \ref{snowline_grainsize} (a) the 3 \micron\ surface height for
the 0.1 \micron-sized grain and 10 \micron-sized grain cases relative to
the 1 \micron-sized grain case. The panel (b) shows those of the
ice-condensation front. From Figure \ref{snowline_grainsize}, we find
that at around 10 AU, the 3 \micron\ surface heights of both of the 0.1
and 10 \micron-sized cases are lower than that of the 1 \micron-sized
case, while the ice-condensation heights of the both cases are similar
to that of the 1 \micron-sized case. This results in a smaller 
$R_{\rm snow}$ for the 0.1 and 10 \micron-sized case than for the 1
\micron-sized case. This behavior is caused by a complex size
dependence of the absorption and scattering opacities with and without
water ice. More studies about this point are required in future.

\section{Summary}

We have investigated the effect of photodesorption on the
ice-condensation front and on the surface snow line position in a
protoplanetary disk around a Herbig Ae/Be star. This study is inspired
by the detection of an absorption feature of water ice at the wavelength
3 \micron\ in the scattered light from the surface of the disk around
HD142527 by \cite{honda09}. We have used 1+1D plane-parallel radiative
transfer algorithm by \cite{dul02} and \cite{inoue09} to obtain the
temperature and density structure in the disk. Then we have determined
the amount of water ice particles by the balance between
photodesorption, sublimation, and condensation, following
\cite{grigorieva07}.

We find that the effect of photodesorption depresses the
ice-condensation front toward the mid-plane and pushes the surface snow
line outward significantly. We also find that the surface snow line
position is determined by the balance between photodesorption and
condensation for a higher effective temperature of the central star,
while it is determined by the balance between sublimation and
condensation for a lower effective temperature. We define the switching
point of the two balances in the effective temperature as the critical
temperature. The surface snow line is located at a small disk radius
($<$ several tens of AU) if the stellar effective temperature is lower
than the critical temperature. In the opposite case, the surface snow
line moves outward rapidly and tends to be at a large disk radius ($>$
several hundreds of AU) due to efficient photodesorption. We also
derive an analytic formula to describe the critical temperature as a
function of stellar and disk parameters.

We discuss a few implications for the observation of the surface snow
line in circumstellar disks. We plot the critical temperature as a
function of the stellar luminosity on the HR diagram. This indicates the
boundary between two mechanisms determining the position of the surface
snow line: the sublimation--condensation balance and the
photodesorption--condensation balance. Most of Herbig Ae/Be stars are
found in the photodesorption area. We also plot the line which gives the
position of the surface snow line of 100 AU on the HR diagram. HD142527
is found slightly above the line, suggesting the surface snow line of
the HD142527 disk is located at 100--300 AU from the central star. This
is consistent with the observations by \cite{honda09} which indicated
the surface snow line was located at $<140$ AU. We also find that there
are many Herbig Ae/Be stars whose snow line on the disk surface is
$>100$ AU or even $>1000$ AU. It is difficult to detect water ice
particles on the surface of the disk around these stars, which will be
easily confirmed by observations. Finally, we also find that the surface
snow line position depends on the dust grain size by a complex way,
which is worth investigating more thoroughly in future.

\acknowledgments

We thank the referee for her/his constructive comments which were
useful to improve the clarity of this manuscript very much.


%% in parentheses, after the keyword, but they will not be verified.



\appendix

\section{Disk structure}

Here, we introduce details of our calculations to determine the density
and temperature structure of a circumstellar disk. Our method is based
on \cite{dul02}, \cite{inoue09}, and \cite{oka11}

\subsection{Hydrostatic equilibrium}

The dynamical structure of the disk in the vertical direction is assumed
to be determined by the balance between the vertical component of the
gravitational force by the central star and the gas pressure gradient
(hydrostatic equilibrium);  
\begin{equation}
 \frac{\partial P(R,Z)}{\partial Z} = -\rho(R,Z)\frac{GM_{*}}{R^{3}}Z,
  \label{eq:hydrostatic_equilibrium}
\end{equation}
where $P(R,Z)$, $\rho(R,Z)$, $G$, and $M_{*}$ are the gas pressure, the
gas volumetric density, the gravitational constant, and the mass of the
central star, respectively. We adopt the equation of state of the ideal
gas for the disk gas; 
\begin{equation}
 P(R,Z)=\frac{\rho(R,Z)}{\mu_{\mathrm{mean}} m_{u}} kT(R,Z),
  \label{eq:state_of_gas}
\end{equation}
where $k$, $T(R,Z)$, $\mu_{\mathrm{mean}}$, and $m_{u}$ are the Boltzmann
constant, the temperature of the disk gas, the mean molecular weight (we
set this value to be 2.3), and the atomic mass unit, respectively.
To obtain the vertical density profile $\rho(R,Z)$ at a radius $R$ by
integrating equation (A1) with (A2) along the vertical axis $Z$, we
need the mid-plane density $\rho(R,0)$ as the normalization of
$\rho(R,Z)$. This is given by equating the surface density $\Sigma(R)$
with the integral of $\rho(R,Z)$ along $Z$ as 
\begin{equation}
 \Sigma(R) = \int_{-\infty}^{\infty} \rho(R,Z) dZ.
  \label{eq:definition_of_surface_density}
\end{equation}
The surface density $\Sigma(R)$ is given by equation (1).

\subsection{Radiative transfer}

The temperature in the disk is determined by the balance between the
cooling and heating. The disk is heated by the irradiation from the
central star. We ignore the heating by the viscous dissipation of the
disk gas, because the snow line at the disk surface is located at
relatively outer part of the disk where the viscous heating is negligibly
small compared with the heating by the irradiation of the central star. 
The cooling rate in the disk is determined by the radiation energy
transfer and the convection is not taken into account because of
the low density in the disk.

Assuming a plane-parallel structure along the $Z$-axis in an annulus at
a certain $R$, the equation of the radiative transfer is given by  
\begin{equation}
 \mu \frac{dI_{\mu,\nu}(Z)}{dZ} = \rho \kappa_{\nu} (S_{\nu}(Z) - I_{\mu,\nu}(Z)),
  \label{eq:radiative_transfer}
\end{equation}
where $\mu$, $\nu$, $I_{\mu, \nu}$, $\kappa_{\nu}$, and $S_{\nu}$ are
the cosine of the angle between the propagation direction and the
$Z$-axis, the frequency, the specific
intensity, the extinction coefficient, and the source function,
respectively. The subscripts $\mu$ and $\nu$ of each quantity
represent the direction cosine and the frequency of that quantity. If we
assume an isotropic scattering and local thermodynamics equilibrium, the
source function $S_{\nu}$ is given by   
\begin{equation}
 S_{\nu}(Z) = (1-\varpi_{\nu}(Z)) B_{\nu}(T) 
  + \varpi_{\nu}(Z) \frac{1}{2}\int_{-1}^{1}I_{\mu,\nu}d\mu 
  + \varpi_{\nu}(Z)\frac{F_{\mathrm{irr},\nu}(Z)}{4\pi}, 
  \label{eq:source_function}
\end{equation}
where $\varpi_{\nu}$, $B_{\nu}(T)$, and $F_{\mathrm{irr},\nu}(Z)$ are
the single scattering albedo, Planck function with temperature $T$, and
radiation flux from the central star, respectively.

We evaluate $F_{\mathrm{irr},\nu}(Z)$ using the so-called grazing angle
recipe (e.g., Chiang and Goldreich 1997, Dullemond et al. 2002) as
\begin{equation}
 F_{\mathrm{irr},\nu}(Z) = \frac{L_{\nu}}{4\pi R^{2}} 
  \exp\left(-\frac{\tau_{\nu}(R,Z)}{\beta} \right),
\end{equation}
where $\beta$, $L_{\nu}$, and $\tau_{\nu}(R,Z)$ are the grazing angle
(i.e. incidence angle of the radiation from the central star into the
disk surface), the luminosity of the central star, and the optical depth
between a point $(R,Z)$ and the infinity along the vertical direction
$(R, \infty)$, respectively. In this study, we set $\beta=0.05$
throughout of the disk for simplicity which is a typical value of the
grazing angle. We assume that the central star emits the black body
radiation with an effective temperature $T_*$. Hence, 
$L_\nu=4\pi^{2} R_{*}^{2}B_{\nu} (T_{*})$, where $R_*$ is the radius of
the photosphere of the central star. The optical depth $\tau_{\nu}$ is
defined as 
\begin{equation}
 \tau_{\nu}(R,Z) = \int_{Z}^{\infty}\kappa_{\nu}\rho(R,Z) dZ.
\end{equation}

The temperature of the dust particles are determined by the assumption
of the radiative equilibrium;  
\begin{equation}
 q_{\mathrm{irr}} + \int_{0}^{\infty}\rho\kappa_{\nu}(1-\varpi_{\nu}) 
  \oint I_{\mu,\nu}(\bm{\Omega})d\Omega d\nu
  = 4\pi \int_{0}^{\infty}\rho\kappa_{\nu}(1-\varpi_{\nu})B_{\nu}(T)d\nu ,
  \label{eq:radiative_equilibrium}
\end{equation}
where $q_{\mathrm{irr}}$ is the heating rate by the irradiation of the
central star and given by 
\begin{equation}
 q_{\mathrm{irr}} = \int_{0}^{\infty}\rho(Z)\kappa_{\nu}
  (1-\varpi_{\nu})F_{\mathrm{irr},\nu}(Z)d\nu.
\end{equation}
We can obtain the diffuse radiation field $I_{\mu, \nu}$, the
temperature of the dust particles $T$, and the source function
$S_{\nu}(T)$ by solving equations (\ref{eq:radiative_transfer}) and
(\ref{eq:radiative_equilibrium}) iteratively. Since a straight forward
way to solve these equations converges very slowly, we employ a variable
Eddington factor method (e.g., Dullemond et al.~2002; Inoue et al.~2009)
for a rapid convergence.

\section{Rates for sublimation, condensation, and photodesorption}

Here, we describe the changing rates of the radius of water ice
particles by sublimation, condensation, and photodesorption. These rates
are according to \cite{grigorieva07}.

The decreasing rate of the radius of water ice particles due to sublimation
$\dot{s}_{\mathrm{sub}}$ is given by
\begin{eqnarray}
 && \dot{s}_{\mathrm{sub}} = -\frac{\dot{m}_{\mathrm{sub}}}
  {4\pi\rho_{\mathrm{ice}} s^{2}} = - \frac{\eta \Phi_{\rm sub}}{\rho_{\mathrm{ice}}} \\
 && \Phi_{\mathrm{sub}} = 4.08 \times 10^{-2}
  \left( \frac{P_{\mathrm{sat}}}{1\mathrm{{torr}}} \right) 
  \left( \mu_{\mathrm{H_{2}O}} \frac{1\mathrm{K}}{T} \right) \mathrm{g\ cm^{-2}s^{-1}},
  \label{eq:sublimation}
\end{eqnarray}
where $\dot{m}_{\mathrm{sub}}$ is the decreasing rate of the grain mass,
$s$ is the grain radius, $\rho_{\mathrm{ice}}$ is the density of water
ice, $\Phi_{\mathrm{sub}}$ is the mass sublimation rate,
$P_{\mathrm{sat}}$ is the saturated water vapor pressure, $\eta$ is the
covering factor (fraction of the surface covered by sublimating
material), and $\mu_{\mathrm{H_{2}O}}$ is the atomic weight of the water
molecule. We set $\rho_{\mathrm{ice}}$ to be 0.92 $\mathrm{g\ cm^{-3}}$
as Miyake \& Nakagawa (1993) and consider that there exists purely icy
dust particles ($\eta = 1$). We adopt the saturated water vapor pressure
same as \cite{grigorieva07}: 
\begin{eqnarray}
 P_{\mathrm{sat}} = 
  \cases{
  2.67 \times 10^{10} \exp \left( - \frac{6141.667\mathrm{K}}{T} \right)\ \mathrm{torr} 
  & ($T \ge 170 \mathrm{K}$),
  \cr
  5.69 \times 10^{12} \exp \left( - \frac{7043.51\mathrm{K}}{T} \right)\ \mathrm{torr} 
  & ($T<170\mathrm{K}$).
  \cr
  }
\end{eqnarray}

The increasing rate of the radius of water ice particles due to
condensation is given by substituting the partial pressure of water
vapor $P_{\mathrm{H_{2}O}}$ for the saturated water vapor pressure
$P_{\mathrm{sat}}$ in equation (\ref{eq:sublimation}). Namely, 
\begin{eqnarray}
 && \dot{s}_{\mathrm{con}} = \frac{\dot{m}_{\mathrm{con}}}
  {4\pi \rho_{\mathrm{ice}} s^{2}} = \frac{\eta \Phi_{\rm con}}{\rho_{\mathrm{ice}}}, \\
 && \Phi_{\mathrm{con}} = 4.08 \times 10^{-2}
  \left( \frac{P_{\mathrm{H_{2}O}}}{1\mathrm{{torr}}} \right) 
  \left( \mu_{\mathrm{H_{2}O}} \frac{1\mathrm{K}}{T} \right) \mathrm{g\ cm^{-2}s^{-1}},
\end{eqnarray}
where $\dot{m}_{\mathrm{con}}$ and $\Phi_{\mathrm{con}}$ are the
increasing rate of the grain mass and the mass condensation rate onto
the grains, respectively. The partial pressure of water vapor
$P_{\mathrm{H_{2}O}}$ is given by the product of the number fraction of
water vapor $X_{\mathrm{H_{2}O}, \mathrm{vapor}}$ and the gas pressure $P$.

The decreasing rate of the radius of water ice particles due to
photodesorption is given by:  
\begin{equation}
 \dot{s}_{\mathrm{pd}} = - \frac{\eta m_{\mathrm{H_{2}O}}
  Y_{\mathrm{pd}} N_{\mathrm{abs}}}{4\rho_{\rm ice}}
\end{equation}
where $m_{\mathrm{H_{2}O}} = 3\times 10^{-23}$ g is the mass of a water
molecule, $Y_{\rm pd}$ is the photodesorption yield, and
$N_{\mathrm{abs}}$ is the number flux of the FUV radiation absorbed by
ice particles. $N_{\mathrm{abs}}(R,Z)$ is given by 
\begin{equation}
 N_{\mathrm{abs}}(R,Z) =
  \int_{\lambda_{\mathrm{min}}}^{\lambda_{\mathrm{max}}}
  \frac{F_\lambda(R,Z)}{hc/\lambda} Q_\lambda^{\rm abs}d\lambda,
  \label{eq:N_abs}
\end{equation}
where $F_\lambda(R,Z)$ is the incident FUV flux of the wavelength
$\lambda$ at the position $(R,Z)$, and $Q_\lambda^{\rm abs}$ is the
absorption coefficient for water ice particles. We set the wavelength
range of the integral of equation (\ref{eq:N_abs}) same as \cite{grigorieva07}: 
$\lambda_{\mathrm{min}} = 0.091~\mathrm{\mu m \ (13.6~eV)}$ and 
$\lambda_{\mathrm{max}} = 0.24~\mathrm{\mu m \ (5.1~eV)}$. 

\cite{westley95a,westley95b} experimentally investigated the
photodesorption of water ice and derived the yield as 
$Y_{\rm pd}\sim1.0\times10^{-3}$. \cite{andersson06} and
\cite{andersson08} carried out a classical molecular dynamics simulation
of photodesorption and derived the yield consistent with
\cite{westley95a,westley95b}. The weakness of their study was that their
calculation was operated with a limited temperature and a limited
wavelength range of the UV radiation. The latest experimental study on
the photodesorption is \cite{oberg09}. They experimentally determined
the rate of the photodesorption of water ice eliminating influences by
other mechanisms and equipments, and obtained more detailed yield as a
function of temperature and the thickness of the molecular layer of
water ice as   
$Y_{\rm pd}(T,\ x>8)=10^{-3}(1.3 + 0.032 T)\ \mathrm{UV\ photon^{-1}}$. 
However, the results of these experimental and theoretical studies about
photodesorption yield still have a large uncertainty. We set 
$Y_{\rm pd}$ to be $1\times10^{-3}$ as \cite{dominik05} and
\cite{grigorieva07} as the fiducial case.

\bigskip

\end{document}